\documentclass[CEJP,PDF]{cej} % use PDF command to enable PDFLaTeX driver
\usepackage{layout}
\usepackage{amsmath}
\usepackage{textcomp}
\usepackage{hyperref}
\usepackage{multirow}
\def\lsim{\raise0.3ex\hbox{$<$\kern-0.75em\raise-1.1ex\hbox{$\sim$}}}
\def\gsim{\raise0.3ex\hbox{$>$\kern-0.75em\raise-1.1ex\hbox{$\sim$}}}

\title{Determination of Freeze-out Conditions from Lattice QCD 
Calculations\footnote{
presented at the International Conference "Critical Point and Onset of Deconfinement - CPOD 2011", Wuhan, November 7-11, 2011;\\
This work has been supported in part by contracts DE-AC02-98CH10886
with the U.S. Department of Energy
and the Bundesministerium f\"ur Bildung und Forschung under grant 06BI9001.}} 
\articletype{Review Article}

\author{Frithjof Karsch\inst{1}$^,$\inst{2}\email{karsch@bnl.gov}
}
\institute{
     \inst{1} Physics Department, Brookhaven National Laboratory,
     Upton, NY 11973, USA
     \inst{2} Fakult\"at f\"ur Physik, Universit\"at Bielefeld,
     D-33615 Bielefeld, Germany
          }

\abstract{Freeze-out conditions in Heavy Ion Collisions are generally
determined by comparing experimental results for ratios of particle
yields with theoretical predictions based on applications of the
Hadron Resonance Gas model. We discuss here how this model dependent
determination of freeze-out parameters may eventually be replaced by
theoretical predictions based on equilibrium QCD thermodynamics.}


\begin{document}
\maketitle

%% ###################################################################

\section{Introduction}

One of the main motivations for the beam energy scan (BES) 
%currently being performed 
at RHIC is to explore the QCD phase diagram at non-vanishing baryon 
chemical potential and to collect evidence for or against the existence of
a critical point at a certain pair ($T,\mu_B$) of temperature ($T$) and 
baryon chemical potential
($\mu_B$) values. Whether or not a phase transition at a parameter set
$(T_{cp},\mu_{cp})$ exists is one of the major uncertainties in our
understanding of the QCD phase diagram.

\begin{figure}[t]
%\begin{center}
%\epsfig{file=phase_3d_freeze_4.eps,width=74mm}
%\end{center}
%\framebox[4cm]{\Huge Figure 1}
\vspace{-8.0cm}
\hspace*{0.2cm}\includegraphics[width=0.5\textwidth]{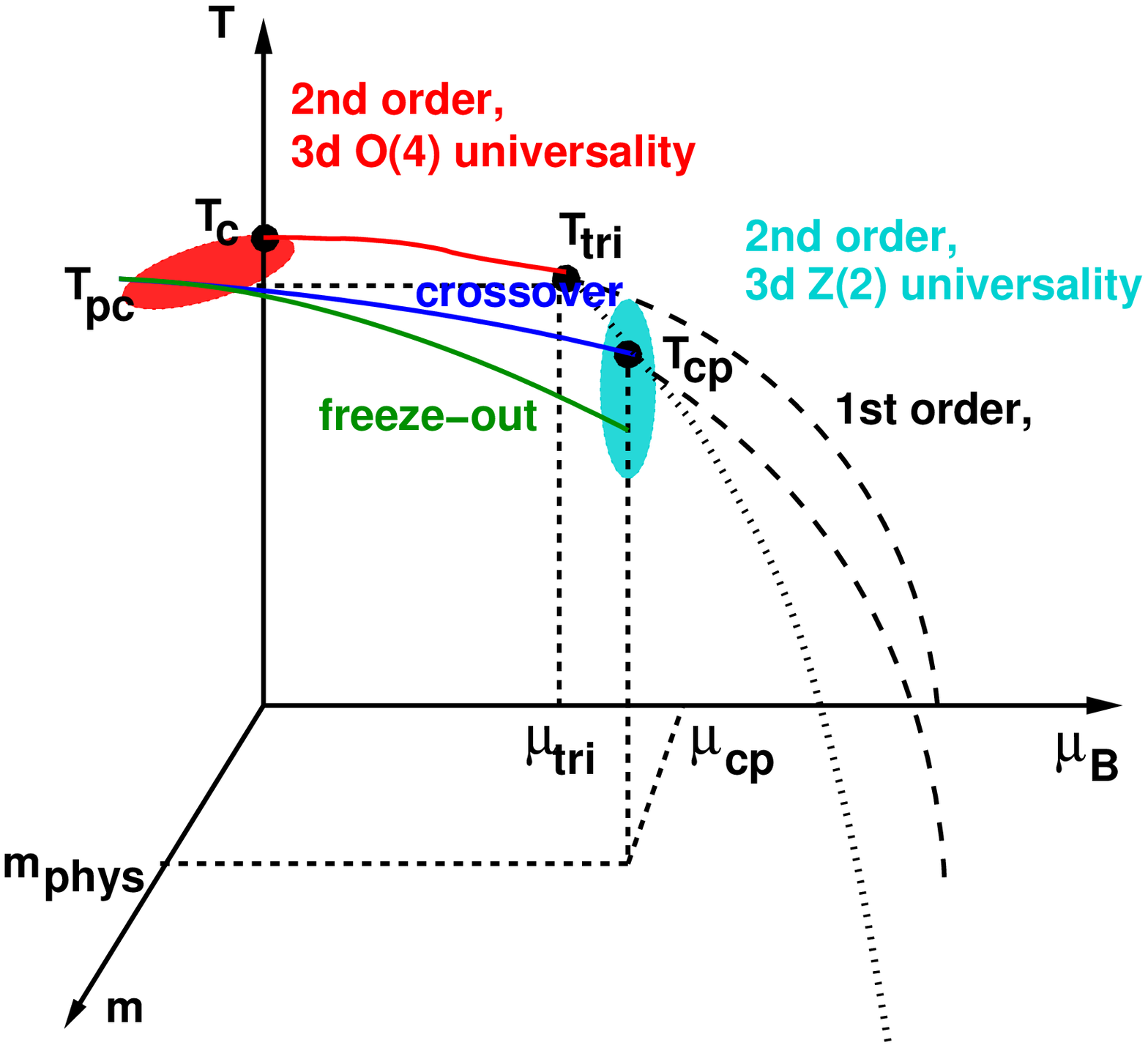}\hspace*{-0.4cm}\includegraphics[width=0.7\textwidth]{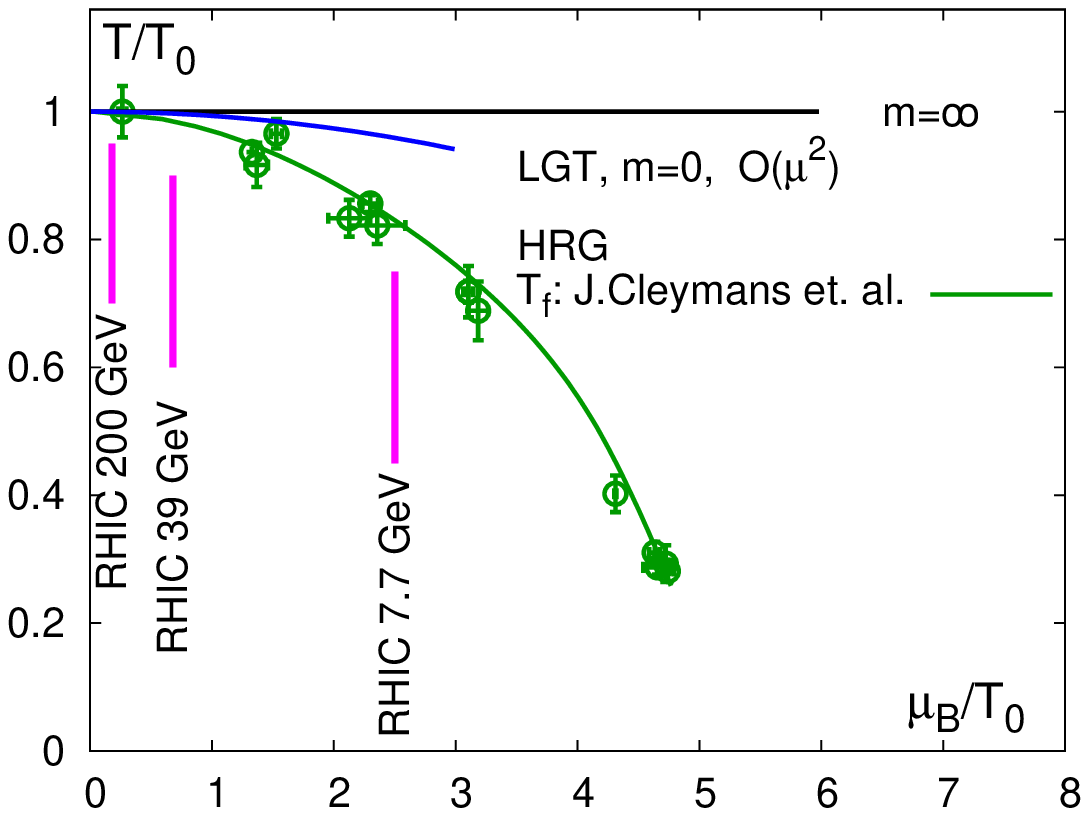}
\vspace*{-0.7cm}
\caption{Phase diagram of QCD in the space of temperature, baryon chemical
potential and light quark mass (left) and the freeze-out line determined
from a comparison of ratios of particle yields and
hadron resonance gas model calculations (right). Also shown in the
right hand figure are results for the chiral phase transition line
calculated in lattice QCD to leading order in the square of the
baryon chemical potential 
\cite{curvature}. 
\vspace*{-0.2cm}
\label{fig:phase}}
\end{figure}

In the vicinity of a critical point various thermodynamic quantities 
will show large fluctuations. However, even if equilibrated,
the hot and dense matter created in a heavy ion collision will expand 
and cool down. Fluctuations of thermodynamic quantities thus, in general
will not be characteristic for a specific ($T,\mu_B$) point in the 
QCD phase diagram. The situation may, however, be different for 
fluctuations of conserved charges that freeze-out at $(T_{f},\mu_{f})$ 
and will not change
afterwards. For this reason the analysis of event-by-event fluctuations
of baryon number, electric charge, and strangeness as well as their higher
order cumulants play a central role in the interpretation of thermal 
conditions created in the BES at RHIC. They provide unique information
about the thermal conditions at the time of chemical freeze-out.
% in a heavy ion collision. 
%If freeze-out occurs in thermal
%equilibrium close to a critical point this should show up in characteristic
%properties of higher order cumulants. 
In fact, this is quite
generally the case and is not only restricted to fluctuations in the 
vicinity of
%of the critical end point at 
$(T_{cp},\mu_{cp})$. It also is
the case at any point on the freeze-out line mapped in the BES. In 
particular, the cumulants of fluctuations of conserved charges will 
also provide information on critical 
behavior at $\mu_B=0$, if the freeze-out points are close 
to the ''true'' chiral phase transition that
exists in QCD for vanishing quark mass values and describes a line 
$T_c(\mu_B)$ in the phase diagram. Whether or not fluctuation observables
will be more sensitive to a possibly existing critical endpoint at
$(T_{cp},\mu_{cp})$ or, for instance, the chiral transition at 
$T_c(\mu_B\simeq 0)$ crucially depends on the proximity of the 
freeze-out parameters $(T_f,\mu_f)$ to the critical region of the 
corresponding critical points. 
In fact, the current determination of freeze-out parameters based 
on Hadron Resonance Gas (HRG) model calculations \cite{cleymans}
and the determination
of the QCD crossover and chiral transition lines, which are known
from lattice calculations in leading order $(\mu_B/T)^2$ 
\cite{curvature,Fodor_curv}, suggest that freeze-out and transition lines
differ more as $\mu_B$ increases. This situation is illustrated in 
Fig.~\ref{fig:phase}.

\section{Freeze-out parameter}

It is common practice in heavy ion phenomenology
to determine the chemical freeze-out parameters
and their dependence on beam energy by comparing experimentally measured
particle yields with HRG model calculations \cite{cleymans}. In fact,
this approach seems to be quite successful and reliable.
It is, however, evident that this approach is conceptually unsatisfactory
and must fail, when freeze-out happens close to a critical point in
the QCD phase diagram where the dependence of thermodynamics on 
$T$ and $\mu_B$
%temperature and baryon chemical potential 
is more complex than in a HRG.  
%A prerequisite for any experimental study of the QCD phase diagram is not
%only that one knows which temperature and chemical potential regime is 
%probed by a measured observable. It clearly also is mandatory to know
%which thermal conditions can be generated in a heavy ion experiment, e.g. 
%by varying the beam energy. When analyzing fluctuations of conserved
%charges, i.e., baryon number, electric charge or strangeness fluctuations,
%the former issue seems to be well controlled -- as long as the system
%generated in a heavy ion collision is in equilibrium, fluctuations of 
%conserved charges probe the thermal conditions at freeze-out. However,
%how are freeze-out temperature and chemical potential(s) determined 
%experimentally? 
Clearly one eventually wants to compare experimental observables
with theoretical predictions based on (equilibrium) QCD. 
%As this is, 
%or was in the past, in practice difficult it has been common practice to
%determine freeze-out conditions by comparing experimental
%results for, e.g. particle yields and suitable ratios of particle yields,
%with model calculations performed in the framework of a statistical
%model, the Hadron Resonance Gas (HRG) model \cite{cleymans}. 
Extracting information on particle yields at finite $T$ 
directly from QCD is difficult, if not impossible.
However, the experimental
measurements of fluctuation observables and their higher order cumulants
\cite{STAR},
which all probe thermal conditions at freeze-out, 
and the improved theoretical calculations of fluctuations of conserved 
charges in lattice regularized equilibrium QCD thermodynamics 
\cite{cheng,Mukherjee}
make it now possible to determine freeze-out
conditions directly from QCD. We will outline in the following a 
determination of $T_f$ and $\mu_f$ at different values of the beam
energy. For simplicity we ignore possible, small non-zero values
of the electric charge and strangeness chemical potentials. 
We also will ignore complications that may arise from the limited 
phase space in which fluctuation observables are being analyzed 
experimentally. Our point here is a conceptual one! 
We present this discussion for the case of baryon number fluctuations
but will later on generalize it to the case of electric
charge fluctuations.

\subsection{The baryon chemical potential at freeze-out}

The $n$-th order cumulants of net baryon number fluctuations, 
$\chi^{B}_{n}$, can be calculated in lattice QCD at vanishing baryon 
chemical potential as suitable derivatives of the pressure $p/T^4$.
For small, non-zero values of $\mu_B$ this allows then to calculate cumulants 
from a Taylor series expansion in $\mu_B/T$,
\begin{equation}
\chi^{B}_{n,\mu} =
\sum_{k=0}^{\infty} \frac{1}{k!}\chi^{B}_{k+n}(T) 
\biggl( {\mu_B \over T}\biggr)^k \;\;\;\;\;\;
{\rm with}\;\;\;\;\;\;
\chi^{B}_{n} =\left. \frac{1}{VT^3}
\frac{\partial^n\ln Z}{\partial(\mu_{B}/T)^n} 
\right|_{\mu_B=0} \; .
\end{equation}
Appropriate ratios of these cumulants are related to
shape parameters of the probability distribution of net baryon
number, i.e., the mean value $M_B$, variance $\sigma_B$, skewness $S_B$ 
and kurtosis $\kappa_B$. In particular, one has

\begin{equation}
\frac{\sigma_B^2}{M_B} = \frac{\chi_{2,\mu}^{B}}{\chi_{1,\mu}^{B}} ,
\;\; ~~~~
S_B\sigma_B = \frac{\chi_{3,\mu}^{B}}{\chi_{2,\mu}^{B}} ,\;\; ~~~~ 
\kappa_B \sigma_B^2 = \frac{\chi_{4,\mu}^{B}}{\chi_{2,\mu}^{B}}  \; .
\end{equation}
Let us consider the Taylor expansion for the simplest even-odd ratio
of cumulants, $\chi^{B}_{2,\mu} / \chi^{B}_{1,\mu}$. In next to 
leading order one finds,

\begin{equation}
\frac{\sigma_B^2}{M_B}\equiv \frac{\chi_{2,\mu}^{B}}{\chi_{1,\mu}^{B}} =
\frac{T}{\mu_{B}} \left[\frac{1+
\frac{1}{2}\frac{\chi_{4}^{B}}{\chi_{2}^{B}}(\mu_B/T)^2
+...}{1+
\frac{1}{6}\frac{\chi_{4}^{B}}{\chi_{2}^{B}}(\mu_B/T)^2+...}
\right] \; .
\end{equation}
A similar relation holds for $\chi^{B}_{3,\mu} / \chi^{B}_{2,\mu}$.
To leading order 
the ratios of even and odd cumulants thus determine directly the ratio
of $\mu_B$ and $T$
%of baryon chemical potential and temperature 
at the time of freeze-out, 
$\sigma_B^2/M_B=(T_f/\mu_f)(1+{\cal O}((\mu_f/T_f)^2))$.
The coefficient of the next-to-leading order correction is small
for all temperatures; current lattice QCD calculations suggest 
$\chi_{4}^{B}/\chi_{2}^{B} < 1.5$ for all temperatures. Therefore, the
systematic errors that arise from ignoring this correction also remains
small for a broad range of beam energies covered in the BES at RHIC.
In fact, the systematic error is at most 2\% at $\sqrt{s_{NN}}=200$~GeV
and rises to about 20\% at $\sqrt{s_{NN}}=39$~GeV. 

{\it \hspace*{1.5cm}Even-odd ratios of cumulants are good observables 
to determine the value of the \\
\hspace*{1.5cm}baryon chemical potential at freeze-out.}

We give results for $\mu_{f}/T_f$ based on measurements of 
$\chi_{2,\mu}^{B}/\chi_{1,\mu}^{B}$ by the STAR collaboration \cite{STAR}
in Table 1. These compare quite well with HRG model calculations.

\begin{table}[t]
\begin{tabular}{|c|c|c|c|}
\hline
\multirow{2}{*}{$\sqrt{s_{NN}}$}&STAR&QCD&HRG\\
\cline{2-4}
~& $\chi_P^{(2)}/\chi_P^{(1)}$& $\mu_f/T_f$&
$\mu_f/T_f$\\
\hline
200 & 5.3(9) & 0.190(30)(4)& 0.183\\
63.4& 2.35(42) & 0.43(8)(3)& 0.43\\
\hline
\end{tabular}
\caption{The ratio of baryon chemical potential and temperature at 
freeze-out determined from measurements of the ratio of squared variance
and mean value of net proton number fluctuations by comparing to lattice
QCD calculations of corresponding cumulants of net baryon number fluctuations
(third column). Results are given for the two largest  values of the beam 
energy scan. The second error in the third column gives an estimate for
the systematic error that arises from neglecting next-to-leading order
corrections in the Taylor expansion (Eq.~(3)). 
The last column gives the result for $\mu_f/T_f$ obtained by
comparing measured particle yields with HRG model calculations.}
\end{table}

\subsection{The freeze-out temperature}
While the ratio of even-odd cumulants is most sensitive to the baryon
chemical potential, the ratio of even-even cumulants is, at leading
order, determined only by $T_f$. For small values
of the baryon chemical potential a low order Taylor series thus again
is sufficient. E.g., one finds
for the ratio of fourth and second order cumulants,
\begin{equation}
\kappa_B \sigma_B^2 \equiv \frac{\chi_{4,\mu}^{B}}{\chi_{2,\mu}^{B}} =
\frac{\chi_{4}^{B}(T)}{\chi_{2}^{B}(T)} \left[\frac{1+
\frac{1}{2}\frac{\chi_{6}^{B}(T)}{\chi_{4}^{B}(T)}(\mu_B/T)^2
+...}{1+
\frac{1}{2}\frac{\chi_{4}^{B}(T)}{\chi_{2}^{B}(T)}(\mu_B/T)^2+...}
\right] \; ,
\end{equation}
where we explicitly point out the $T$-dependence of cumulants at $\mu_B=0$.
A potential difficulty in the determination of the freeze-out temperature 
from measurements of $\kappa_B\sigma_B^2=\chi_{4,\mu}^{B}/ \chi_{2,\mu}^{B}$ 
is that lattice QCD calculations \cite{cheng} suggest that 
this quantity varies rapidly only in the crossover region but shows
little variation at low temperature where it stays close to unity. The
discretization errors inherent in these calculations are, however, still
too large to allow a direct comparison with experimental data. This will
change when improved calculations with a better fermion discretization
scheme and closer to the continuum limit will be completed 
\cite{Mukherjee}. 

\section{Freeze-out conditions from electric charge fluctuations}

The discussion presented in the previous section carries over to fluctuations
of other conserved charges, e.g. electric charge or strangeness. The former
is of particular interest, as it may soon be accessible experimentally.
It will also avoid the problems that arise from the fact that the conserved
net baryon number is not accessible directly in a heavy ion experiment. What
is measured instead is the fluctuation of proton number, which may change 
even after freeze-out \cite{Asakawa}. Also theoretically, electric charge
fluctuations allow a more concise determination of freeze-out parameters
as  $\kappa_Q\sigma_Q^2$ shows a characteristic variation with temperature 
also in the hadronic phase \cite{cheng}. 

{\it \hspace*{1.5cm}A measurement of $\chi_{4,\mu}^{Q}/ \chi_{2,\mu}^{Q}$ will
allow to determine the freeze-out temperature.}

More precisely, it determines a line $T_f(\mu_B)$ in the QCD phase diagram;
the additional 
measurement of an even-odd cumulant ratio will then fix $\mu_B\equiv\mu_f$
and calculations of further ratios will provide consistency checks.
% within equilibrium QCD.

\begin{figure}[t]
\vspace{-8.0cm}
\includegraphics[width=0.7\textwidth]{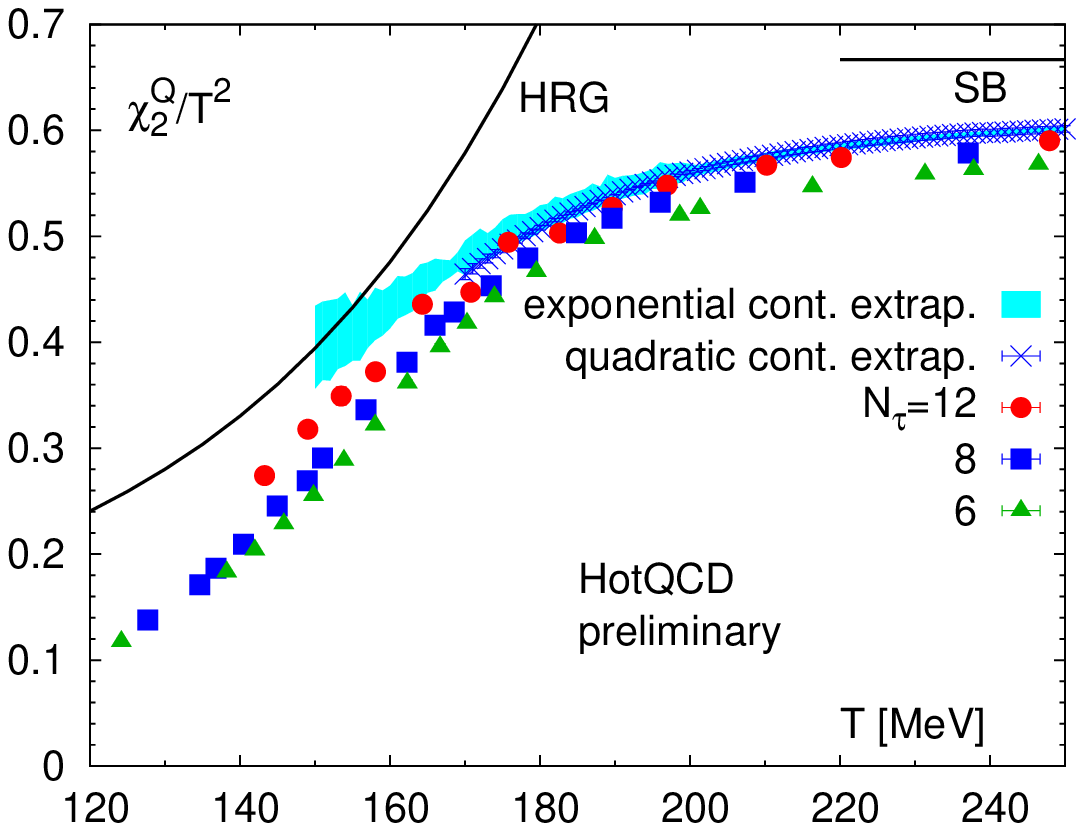}\hspace*{-4.0cm}
\includegraphics[width=0.7\textwidth]{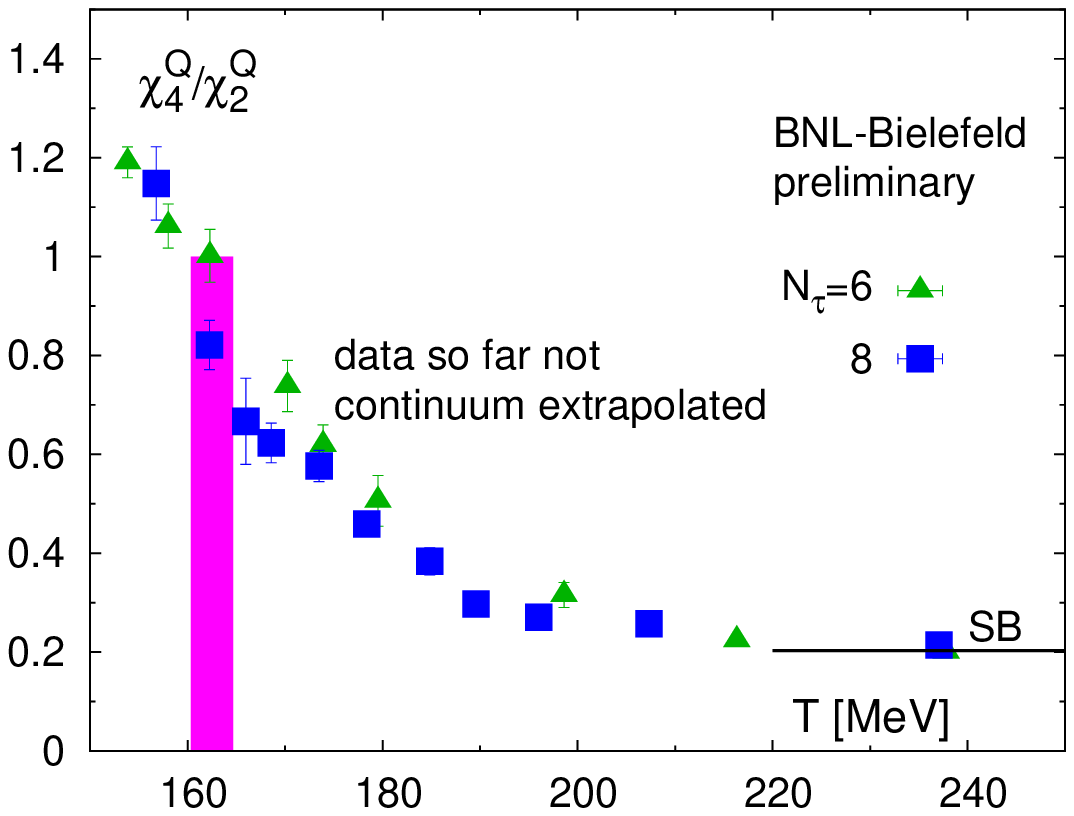}
\vspace{-1.5cm}
\caption{Quadratic fluctuations of net electric charge at $\mu_B=0$
(left) \cite{hotQCD} and preliminary results for the ratio of
quartic to quadratic fluctuations \cite{BNLBI}. Results are compared 
to HRG model calculations at low temperatures (HRG) and an ideal 
quark gas (SB) at high temperature.
For comparison, in a HRG at $T\simeq 160$~MeV, one has 
$\chi_4^Q/\chi_2^Q\simeq 1.7$.
%The difference of the curve labeled "HRG-($m_{RMS}$)" and the "HRG" curve
%indicates the size of cut-off effects in current determinations of
%this ratio from lattice QCD calculations at low temperature.
\label{fig:chi42}}
\end{figure}

In Fig.~\ref{fig:chi42}(left) we show results for the quadratic fluctuations
of net electric charge and compare this to HRG model calculations 
\cite{hotQCD}. 
Preliminary results for $\chi_{4,\mu}^{Q}/ \chi_{2,\mu}^{Q}$ \cite{BNLBI}
are shown in Fig.~\ref{fig:chi42}(right). We stress that the latter require
a careful cut-off analysis, which already for quadratic electric charge
fluctuations is difficult \cite{hotQCD}.

Fig.~\ref{fig:chi42} shows that QCD results for 
quadratic electric charge fluctuations
are consistent with HRG model calculations only for 
temperatures $T\lsim 160$~MeV. Preliminary results for the quartic
fluctuations \cite{BNLBI} suggest that the ratio 
$\chi_{4}^{Q}/ \chi_{2}^{Q}$ 
differs strongly from HRG results at $T\simeq 160$~MeV, i.e.
$\chi_{4}^{Q}/ \chi_{2}^{Q})_{HRG}\simeq 1.7$, and is less than
unity.

\section{Conclusion}
Even-odd ratios of cumulants of conserved charge fluctuations allow to
determine the baryon chemical potential at freeze-out.
A more precise experimental determination of 
$\kappa_B\sigma_B^2$ as well as $\kappa_Q\sigma_Q^2$
and improved lattice QCD results for these observables
will not only allow to determine the freeze-out temperature $T_f$, it
will also provide information on the deviation of the freeze-out line
$T_f(\mu_B)$  from the 
crossover line $T_{pc}(\mu_B)$ and the chiral phase transition
line $T_{c}(\mu_B)$. 

\begin{thebibliography}{99}
\bibitem{cleymans}
J.~Cleymans, H.~Oeschler, K.~Redlich and S.~Wheaton,
  %``Comparison of chemical freeze-out criteria in heavy-ion collisions,''
  Phys.\ Rev.\  C {\bf 73}, 034905 (2006).
\bibitem{curvature}
O.~Kaczmarek {\it et al.},
%F.~Karsch, E.~Laermann, C.~Miao, S.~Mukherjee, P.~Petreczky, C.~Schmidt, W.~Soeldner {\it et al.},
  %``Phase boundary for the chiral transition in (2+1) -flavor QCD at small values of the chemical potential,''
  Phys.\ Rev.\  {\bf D83}, 014504 (2011).
\bibitem{Fodor_curv}
  G.~Endrodi, Z.~Fodor, S.~D.~Katz and K.~K.~Szabo,
  %``The QCD phase diagram at nonzero quark density,''
  JHEP {\bf 1104}, 001 (2011),
  [arXiv:1102.1356 [hep-lat]]
\bibitem{STAR}
M.~M.~Aggarwal {\it et al.} (STAR Collaboration), Phys. Rev. Lett. {\bf 105},
22302 (2010).
\bibitem{cheng}
M.~Cheng
%P.~Hendge, C.~Jung, F.~Karsch, O.~Kaczmarek, E.~Laermann, R.~D.~Mawhinney, C.~Miao 
{\it et al.},
  %``Baryon Number, Strangeness and Electric Charge Fluctuations in QCD at High Temperature,''
  Phys.\ Rev.\  {\bf D79}, 074505 (2009).
\bibitem{Mukherjee}
for a recent overview see: S.~Mukherjee,
  %``Fluctuations, correlations and some other recent results from lattice QCD,''
  J.\ Phys.\ G G {\bf 38}, 124022 (2011),
  [arXiv:1107.0765 [nucl-th]].
%\bibitem{Hegde}
%P.~Hegde (for the HotQCD Collaboration),
%  %``Fluctuations and Higher Moments of Conserved Charges from the Lattice,''
%  arXiv:1110.5932 [hep-lat].
\bibitem{hotQCD}
A. Bazavov {\it et al.} (HotQCD Collaboration), in preparation.
\bibitem{BNLBI}
BNL-Bielefeld Collaboration, in preparation.
\bibitem{Asakawa}
M.~Kitazawa and M.~Asakawa,
%``Revealing baryon number fluctuations from proton number fluctuations in relativistic heavy ion collisions,''
  arXiv:1107.2755 [nucl-th].
\end{thebibliography}
\end{document}